# *In vivo* evidence of alternative loop geometries in DNA-protein complexes

Leonor Saiz* and Jose M.G. Vilar*

* Integrative Biological Modeling Laboratory, Computational Biology Program, Memorial Sloan-Kettering Cancer Center, 1275 York Avenue, Box #460, New York, NY 10021, USA

ABSTRACT   The *in vivo* free energy of looping double-stranded DNA by the *lac* repressor has a remarkable behavior whose origins are not fully understood. In addition to the intrinsic periodicity of the DNA double helix, the *in vivo* free energy has an oscillatory component of about half the helical period and oscillates asymmetrically with an amplitude significantly smaller than predicted by current theories. Here, we show that the *in vivo* behavior is accurately accounted for by the simultaneous presence of two distinct conformations of looped DNA. Our analysis reveals that these two conformations have different optimal free energies and phases and that they behave distinctly in the presence of key architectural proteins.

The formation of DNA loops by the binding of proteins at distal DNA sites plays a fundamental role in many cellular processes, including transcription, recombination, and replication (4-7). In the regulation of gene expression, proteins bound far away from the genes they control can be brought to the initiation of transcription region by looping the intervening DNA. The free energy cost of this process determines how easily DNA can form loops and therefore the extent to which distal DNA sites can affect each other (7). Assessing directly the *in vivo* value of the free energy of DNA looping is remarkably difficult. The cell is a densely packed dynamic structure made of thousands of different molecular species that strongly interact with each other. Such complexity poses strong barriers for experimentally characterizing the cellular components, not only because the properties of the components can change when studied *in vitro*, but also because the *in vivo* probing of the cell can perturb the process under study (8,9).

Computational modeling was recently used to infer the *in vivo* free energies of DNA looping by the *lac* repressor (10) as a function of the loop length (1) from measurements of enzyme production in the *lac* operon (2). This analysis showed that the free energy for short loops oscillates with the helical periodicity of DNA, as expected, because the operators must have the right phase to bind simultaneously to the repressor and, unexpectedly, that the free energy in a cycle behaves asymmetrically. A Fourier analysis of the oscillations indicated that this asymmetry can be characterized by a second representative oscillatory component with a period of ~5.6 bp in addition to the component with the *in vivo* helical period (~10.9 bp). Another striking feature of the *in vivo* free energy is that the amplitude of the oscillations is ~2.5 kcal/mol, similar to the typical free energy of cooperative interactions between regulatory molecules (11).

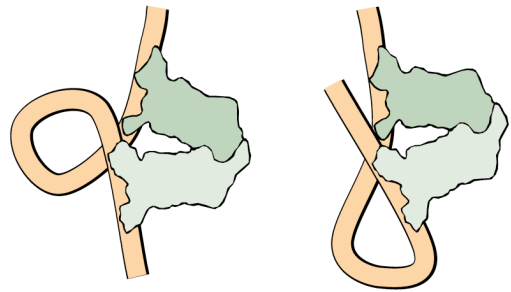

**FIGURE 1** Two alternative loop conformations of the *lac* repressor-DNA complex: the bidentate repressor, with the two dimers that form the functional tetramer colored with different green shades, simultaneously binds the DNA, colored orange, at two sites.

Uncovering the origin of these novel properties is important for understanding DNA looping and its effects in gene regulation, especially because current theories predict symmetric and, at least, twice as big oscillations (12-14). Different contributions, such as the anisotropic flexibility of DNA, local features resulting from the DNA sequence (15), and interactions with the *lac* repressor (16) and other DNA binding proteins, might be at play.

Depending on the orientation of the two DNA binding sites and the properties of the looped DNA-protein complex, the DNA loop can be accomplished following different trajectories (17-19). Thus, the observed behavior could be the result of loops with several representative configurations (Figure 1). Here we show that the oscillatory *in vivo* behavior of the free energy for short loops can be accurately accounted for by the simultaneous presence of two distinct types of DNA loops with different optimal free energies and phases.

If the DNA loop can be in two distinct representative conformations, the free energy of looping, $\Delta G_l$, can be

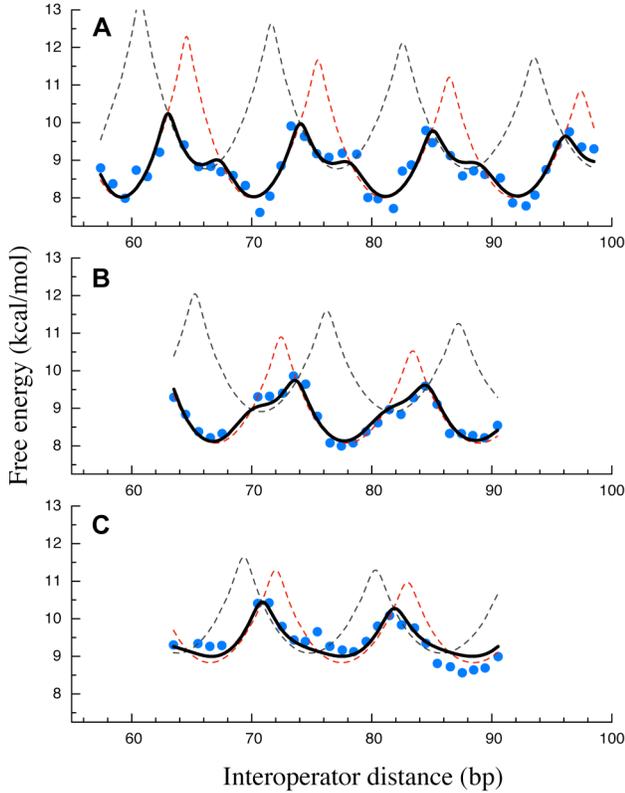

**FIGURE 2** *In vivo* free energy of looping DNA by the *lac* repressor (symbols) obtained as described in Saiz et al. (1) from the measured repression levels of Muller et al. (2) for wild type (A) and of Becker et al. (3) for wild type (B) and a mutant that does not express the architectural HU protein (C). As repression levels in the absence of looping we have used (A) 135, (B) 2.3, and (C) 1.7. The thick black lines are the best fit to the free energy $\Delta G_l$ given by Equation 1, which considers the contributions of two loop conformations. The contributions of each conformation are shown separately as red ($\Delta G_{0,1}$) and gray ($\Delta G_{0,2}$) dashed lines. The values of the parameters for the best fit are shown in Table 1.

expressed as the average over the free energy of each conformation:

$$\Delta G_l = \sum_{i=1}^{2} \sum_{n=-\infty}^{\infty} \frac{1}{Z} \Delta G_{i,n} e^{-\Delta G_{i,n}/RT}, \quad (1)$$

where the index $i$ indicates whether the loop is in the configuration labeled 1 or 2, $Z = \sum_{i=1}^{2} \sum_{n=-\infty}^{\infty} e^{-\Delta G_{i,n}/RT}$ is the normalization factor, and $RT$ (=0.6 kcal/mol) is the gas constant, $R$, times the absolute temperature, $T$.

The free energy of a particular representative conformation includes bending and twisting contributions and is given following the elasticity theory of DNA (12) by:

$$\Delta G_{i,n} = \Delta G_{0,i} + \frac{C}{2L} \frac{4\pi^2}{hr^2} \left( L - L_{opt,i} + n \cdot hr \right)^2, \quad (2)$$

where $L$ is the length of the loop (in bp), $L_{opt,i}$ is the optimal spacing or phase (in bp), and $\Delta G_{0,i}$ is the corresponding optimal free energy (in kcal/mol), which depends on the type of loop formed. The twisting force constant (in kcal/mol bp), $C$, and the *in vivo* helical repeat (in bp), $hr$, are considered to be the same for the two types of loops. The integer index $n$ ranges from -infinity to +infinity and accounts for the $2\pi$ degeneracy in the twisting angle. Note that in the absence of twisting, the free energy of a loop conformation is just $\Delta G_{0,i}$.

We have used this approach to analyze the *in vivo* free energies of looping DNA (1) obtained from the measured repression levels for two wild type situations (2,3) and a mutant lacking the architectural HU (heat unstable nucleoid) protein (3). The free energy of looping $\Delta G_l$ given by Equations 1 and 2 closely reproduces the broad range of observed behaviors (Figure 2), which include not only asymmetric oscillations with reduced amplitude but also plateaus and secondary maxima. The type of behavior depends on the properties of the different loop conformations (Table 1).

In both of the wild type situations analyzed, the presence of two loop conformations (one more stable than the other by 0.8 kcal/mol and with a shift in the optimal phase of 3.9 bp or -3.9 bp) is responsible for the reduced amplitude of the oscillations and the asymmetry of the curves. As the distance between the two operators is changed, the less stable loop becomes the most stable one. Thus, alternative loop conformations are adopted as the length of the loop is changed.

**TABLE 1** Best fit values of the parameters of the free energy of looping expression with two distinct loop conformations (Equations 1 and 2) for the free energies obtained from Muller et al. (2) for wild type (A) and from Becker et al. (3) for wild type (B) and a mutant that does not express the architectural HU protein (C).

| Case | $\Delta G_{0,1}$ (kcal/mol) | $\Delta G_{0,2}$ (kcal/mol) | $L_{opt,1}$ (bp) | $L_{opt,2}$ (bp) | $hr$ (bp) | $C$ (kcal/mol bp) |
|------|------|------|------|------|------|------|
| A | 8.0 | 8.8 | 4.4 | 0.5 | 10.9 | 56 |
| B | 8.1 | 8.9 | 1.0 | 4.9 | 11.0 | 41 |
| C | 8.8 | 9.1 | 0.7 | -2.0 | 11.0 | 35 |

In the mutant without architectural HU proteins, the *in vivo* free energy of DNA looping is compatible with the presence of two loop conformations that are similarly stable (0.3 kcal/mol difference) but have different optimal phases. In this case, the phase shift (2.7 bp) also leads to a reduced amplitude of the oscillations, as in the wild-type case, yet the asymmetric behavior is practically lost; now the presence of two loop conformations results in symmetric oscillations with smaller amplitude.

In all three cases studied, the results obtained for the apparent *in vivo* twisting force constants (in the range 35-56 kcal/mol bp), which also include the contributions from the repressor, are at least a factor 2 smaller than the canonical values (105 kcal/mol bp) (12), and are similar to those reported in Ref. (20). Thus, it is relatively easy for proteins in the intracellular environment to circumvent the constrains that twisting imposes to the formation of DNA loops.

The mathematical expression for the free energy of looping (Equations 1 and 2) indicates that the asymmetric behavior is the result of the presence of a stabilized loop conformation, as illustrated in Fig. 2. There is only the particular exception of a phase difference of exactly 0.5 times the helical period, which would lead to symmetric curves. Symmetric profiles are, in general, consequence of the presence of equally stable loop conformations. Our analysis indicates that in *E. Coli* cells, HU proteins stabilize preferentially one loop conformation and lead to the observed asymmetry. Interestingly, different loop trajectories have been observed for different types of nucleoprotein complexes that loop DNA and in the presence of key architectural proteins analogous to the HU protein (17).

Our analysis has revealed that the formation of DNA loops *in vivo* is tightly coupled to the molecular properties of the proteins and protein complexes that form the loop. There is a high versatility of looped DNA-protein complexes at establishing different conformations in the intracellular environment and at adapting from one conformation to another. This versatility underlies the unanticipated behavior of the *in vivo* free energy of DNA looping and can be responsible not only for asymmetric oscillations with decreased amplitude but also for plateaus and secondary maxima. All these features indicate that the physical properties of DNA can actively be selected to control the cooperative binding of regulatory proteins and to achieve different cellular behaviors.

## REFERENCES AND FOOTNOTES